\documentclass[prd,onecolumn,nofootinbib]{revtex4}

\usepackage{amsmath}
\usepackage{graphicx}
\usepackage{amsfonts}
\usepackage{latexsym}
\usepackage{bbold}
\usepackage{wasysym}
\usepackage{calligra}
\usepackage{float}
\usepackage{ulem}
\usepackage{inputenc}
\usepackage{xspace}
\usepackage{url}
\usepackage{epstopdf}
\usepackage{tikz}
\usepackage{amssymb}
\usepackage{enumitem}

\usepackage{appendix}

\usepackage{amsmath}
\usepackage{amssymb}
\usepackage{graphicx}
\usepackage{amsfonts}
\usepackage{latexsym}
\usepackage{bbold}
\usepackage{wasysym}
\usepackage{calligra}
\usepackage{float}
\usepackage{ulem}
\usepackage{inputenc}
\usepackage{xspace}
\usepackage{url}
\usepackage{epstopdf}
\usepackage{tikz}
\usepackage{amsthm}
\usepackage{soul}

\newcommand{\be}{\begin{equation}}
\newcommand{\ee}{\end{equation}}
\newcommand{\beq} {\begin{equation}}
\newcommand{\eeq} {\end{equation}}
\newcommand{\ba}{\begin{eqnarray}}
\newcommand{\ea}{\end{eqnarray}}

\newcommand{\tdnb}{\tilde{\nabla}}

\begin{document}

	\title{Non-local Metric-Affine Gravity}
	
	\author{Salvatore Capozziello
    \textsuperscript{1,2,3}}
    \author{Damianos Iosifidis\textsuperscript{2,3}}  
    \affiliation{$^1$ Dipartimento di Fisica "E. Pancini", Universit\'a di Napoli "Federico II", Complesso Universitario di Monte S. Angelo, Via Cinthia Edificio 6, I-80126 Napoli, Italy}    \affiliation{$^{2}$Scuola Superiore Meridionale, Largo San Marcellino 10, 80138 Napoli, Italy}
    \affiliation{$^{3}$Istituto Nazionale di Fisica Nucleare – Sezione di Napoli, Via Cinthia, 80126 Napoli, Italy} 
\email{capozziello@na.infn.it, d.iosifidis@ssmeridionale.it}	
	
	\date{\today}
	\begin{abstract}
    Non-local  gravity can potentially solve several problems of  gravitational field both at Ultra-Violet and Infra-Red scales. However, such an approach has been formulated mainly in metric formalism. In this paper, we discuss non-local theories of gravity in the metric-affine framework. In particular, we study the dynamics of  metric-affine analogue of some  well-studied non-local theories, by treating the metric and the connection as independent fields. The approach  gives the opportunity to deal with non-local gravity under a more general standard. Furthermore, we  introduce some novel non-local metric-affine theories with no Riemannian analogue and investigate their dynamics. Finally we discuss some   cosmological applications of our development.
		
	\end{abstract}
	
	\maketitle
	
	\allowdisplaybreaks
	
	

\section{Introduction}

It is widely believed that quantum gravitational effects manifest through non-locality. This feature is invoked in order to reconcile General Relativity (GR)  with Quantum Field Theory (QFT) prescriptions. These quantum   corrections, that alter the classical gravitational action, encode  non-locality and dramatically change the associated dynamics producing new degrees of freedom.
Non-locality is typically expressed through the inclusion of inverse powers of the box operator $\Box^{-1}$  sandwiched between geometric  quantities (like the Ricci scalar) that supplement the gravitational action and consequently the associated field equations \cite{Mashhoon:2017qyw, Deser:2007jk,Capozziello:2008gu,Nojiri:2010pw}. Another kind of non-locality is realized by adding  an infinite series of the box operator in the form of entire functions like e.g. $R F(\Box)R$ where $R$ is the Ricci scalar and $F(\Box)$ a holomorphic function of the box operator. This is known as Infinite Derivative Gravity (IDG) \cite{efimov1967non}. Besides fundamental physics issues, the key idea to adopt non-local gravity models is to obtain characteristic lengths and masses capable of addressing Ultra-Violet and Infra-Red issues in astrophysics and cosmology \cite{Blome:2010xn, Roshan:2022zov,Boos:2020kgj,Boos:2018bxf}.

Cosmological applications  of non-local gravity have been studied in Refs.  \cite{Deser:2007jk, Koivisto:2008xfa, Calcagni:2018pro, Capozziello:2024qol,Capozziello:2021krv,Nojiri:2019dio, Biswas:2010zk, Kolar:2021qox, Koshelev:2016xqb, Calcagni:2013vra, Dirian:2014bma} and some problems like the Weinberg no-go theorem for cosmological constant can be addressed in this framework \cite{Capozziello:2025bsm}. Non-local hybrid metric-Palatini cosmology was studied in \cite{Bombacigno:2024lud}. The teleparallel formulation of non-local gravity along with its cosmological applications, has been considered in \cite{Bahamonde:2017bps}. Also black hole solutions can be improved with the presence of non-local terms \cite{Bambi:2016uda, Capozziello:2025iwn} as well as gravitational waves where new polarizations emerge \cite{Capozziello:2024bjb,Foffa:2013sma}.

Here we are going to formulate non-local gravity in the  metric-affine geometry setup in order to consider the most general framework where these theories can be formulated. This approach proves to be useful also in view of extensions and modifications of General Relativity where the geometric invariants can be other than the standard curvature \cite{Bahamonde:2017bps,BeltranJimenez:2019esp, Capozziello:2022zzh, Capozziello:2025hyw, Golovnev:2023yla}.

The paper is organized as follows. In Sec.$II$, we introduce the notation and conventions we are going to use. Then, in Sec. $III$,  we start with the metric-affine analogue of the  theory introduced in Ref. \cite{Gottlober:1989ww}, which was studied  in the Riemannian setting. Immediately the differences with the metric formalism emerge. By solving the connection field equations of the theory we cast it in an equivalent Scalar-Tensor metric theory with higher order derivative terms for the scalar field. Consequently we generalize it in Sec. $IV$ and study the metric-affine version of the general class of theories with the Lagrangians $\mathcal{L}=R F(\Box)R$ and  $\mathcal{L}=R F(\Box^{-1})R$. Deriving  the field equations, and solving the connection equations, we again establish equivalence with a generalized higher-derivative Scalar-Tensor theory over a Riemannian background. Finally, we introduce a new non-local Metric-Affine Gravity theory which has no Riemannian analogue and consists of the newly introduced terms $X\Box X$, where $X$ stands for torsion and non-metricity, into the action, along with the usual Einstein-Hilbert term. Obtaining the field equations we show how torsion and non-metricity are dynamical fields in this setting even in vacuum.  Some applications to cosmology are discussed in Sec $VI$. In Sec. $VII$, we then wrap up our results and point to future directions.

\section{The Metric-Affine formalism}

 Let us consider a $n$-dim manifold equipped with a metric $g$ and a linear connection $\nabla$, whose components, in local coordinates read $g_{\mu\nu}$ and $\Gamma^{\lambda}{}_{\mu\nu}$ respectively. With these at hand, we can accordingly define curvature, torsion and non-metricity, respectively, as
 \begin{subequations}
\label{eq: spin only}
\begin{align}
R^{\mu}_{\;\;\;\nu\alpha\beta}&:= 2\partial_{[\alpha}\Gamma^{\mu}_{\;\;\;|\nu|\beta]}+2\Gamma^{\mu}_{\;\;\;\rho[\alpha}\Gamma^{\rho}_{\;\;\;|\nu|\beta]} \;\;,\label{R} \\
S_{\mu\nu}^{\;\;\;\lambda}&:=\Gamma^{\lambda}_{\;\;\;[\mu\nu]}
\;\;, \;\;  \\
Q_{\alpha\mu\nu}&:=- \nabla_{\alpha}g_{\mu\nu}
\end{align}
\end{subequations}
The Ricci tensor for the general connection is given by the contraction
\beq
R_{\mu\nu}:=R^{\alpha}_{\;\;\;\mu\alpha\nu}
\eeq
It should be noted that this is not symmetric in general. Subsequent contraction with the metric defines the generalized scalar curvature
\beq
R:=g^{\mu\nu}R_{\mu\nu}
\eeq
Now, from torsion and non-metricity, one can construct the three vectors
\beq
S_{\mu}:=S_{\mu\nu}{}{}^{\nu}\;\;, \;\; Q_{\mu}:=Q_{\mu\alpha\beta}g^{\alpha\beta}\;\;, \;\; q_{\mu}:=Q_{\alpha\beta\mu}g^{\alpha\beta}
\eeq
The difference between the general connection $\Gamma^{\lambda}{}_{\mu\nu}$ and the Levi-Civita one, $\tilde{\Gamma}^{\lambda}{}_{\mu\nu}$,  defines a tensor that quantifies how much the geometry deviates from being Riemannian. This is the so-called distortion tensor \cite{schouten1954ricci,hehl1995metric}

\beq
N^{\lambda}{}_{\mu\nu}:=\Gamma^{\lambda}{}_{\mu\nu}-\tilde{\Gamma}^{\lambda}{}_{\mu\nu} \label{decomposition}
\eeq
Then, using the above definitions of torsion and non-metricity, it is a simple matter to show that the general affine-connection may be decomposed according to
\begin{gather}
\Gamma^{\lambda}_{\;\;\;\;\mu\nu}=\tilde{\Gamma}^{\lambda}_{\;\;\;\mu\nu}+
\frac{1}{2}g^{\alpha\lambda}(Q_{\mu\nu\alpha}+Q_{\nu\alpha\mu}-Q_{\alpha\mu\nu}) -g^{\alpha\lambda}(S_{\alpha\mu\nu}+S_{\alpha\nu\mu}-S_{\mu\nu\alpha}) \label{N}
\end{gather}
This is called the post-Riemannian expansion of the connection and it is particularly useful for splitting the various quantities into their Riemannian parts\footnote{Namely the ones computed with respect to the Levi-Civita connection.} plus the non-Riemannian contributions of torsion and non-metricity. Torsion and non-metricity can then be recovered from the distortion from the relations (see e.g. \cite{Hehl:1994ue}),
\beq
Q_{\mu\alpha\beta}=2 N_{(\alpha\beta)\mu} \;\;, \;\; S_{\mu\nu\alpha}= N_{\alpha [\mu\nu]}. \label{QSN}
\eeq

  Our definition of the box operator is the usual one that involves only the metric, i.e.
\beq
\Box f:=\frac{1}{\sqrt{-g}}\partial_{\mu}(\sqrt{-g}\partial^{\mu}f) \label{Boxdef}
\eeq
on a test function $f$. In a covariant manner, it can also be written as 
\beq
\Box=g^{\mu\nu}\widetilde{\nabla}_{\mu}\widetilde{\nabla}_{\nu}=\widetilde{\nabla}^{\mu}\widetilde{\nabla}_{\mu}\,.
\eeq
In this form, it can further  act upon tensor fields of arbitrary rank, $\Box T^{\mu_{1}...\mu_{p}}{}{}_{\nu_{1}...\nu_{q}}=\widetilde{\nabla}^{\mu}\widetilde{\nabla}_{\mu}T^{\mu_{1}...\mu_{p}}{}{}_{\nu_{1}...\nu_{q}}$. Note that in \cite{Briscese:2015urz} the definition 
\beq
\Box^{(\Gamma)}=g^{\mu\nu}\nabla_{\mu}\nabla_{\nu}, \label{Boxga}
\eeq
where $\nabla$ is the full covariant derivative associated with the full connection $\Gamma^{\lambda}{}_{\mu\nu}$, was used in a Palatini formulation of non-local gravity. Although this could be a viable option, there are several reasons to avoid such a definition in the current context. First and foremost, (\ref{Boxga}) does not reduce to (\ref{Boxdef}) when acted upon functions. Secondly, given the fact that the metric is not covariantly constant with respect to the full connection, the above definition is not the only one that reduces to $\partial^{\mu}\partial_{\mu}$ in the flat space limit. Indeed, operators like $\nabla_{\mu}(g^{\mu\nu}\nabla_{\nu})$,  $\frac{1}{\sqrt{-g}}\nabla_{\mu}(\sqrt{-g}g^{\mu\nu}\nabla_{\nu})$ or $\nabla_{\mu}(\nabla_{\nu}g^{\mu\nu})$ also reduce to $\partial^{\mu}\partial_{\nu}$ in the flat space limit. One would then need to take all possible linear combinations of such operators (with the coefficients added up to one) in order not to discriminate among them, see for instance \cite{Shimada:2018lnm}. However, this approach, not only does it    complicate the analysis but also introduces new arbitrary parameters\footnote{\textcolor{black}{As an example, let us consider the linear combination $\Box=a g^{\mu\nu}\nabla_{\mu}\nabla_{\nu}+b\nabla_{\mu}g^{\mu\nu}\nabla_{\nu}$. Consistency in the flat space limit then demands that $a+b=1$. Furthermore using the connection decomposition (\ref{decomposition}) it follows that $\Box f=\Box_{g}f+\Big( \frac{Q^{\mu}}{2}+2S^{\mu}+(b-1)q^{\mu}\Big) \partial_{\mu}f$. That is, it can always be split into the Riemannian box plus derivative couplings of the function and the non-Riemannian vectors.}}. Thirdly, for operators like (\ref{Boxga}), terms like $R_{\mu\nu}\Box^{(\Gamma)} R^{\mu\nu}$ and $R^{\mu\nu}\Box^{(\Gamma)} R_{\mu\nu}$ do not coincide since the metric cannot freely move through the full covariant derivative $\nabla_{\mu}$. This then complicates the analysis even further since one would be forced to include both of them in order to maintain full generality. Finally, the basic property of the inner product $(\phi \Box \psi)=(\Box \phi, \psi)$ is lost if one uses (\ref{Boxga}) instead of (\ref{Boxdef}). It is therefore only natural to stick to (\ref{Boxdef}) for the definition of the box operator in our current context.\footnote{The same definition of the box operator was also used in \cite{Bombacigno:2024lud}.}

\section{A simple Warm-up Theory}

Let us start with the  model in \cite{Gottlober:1989ww}, but now in the metric-affine formalism, where the connection and the metric are independent fields. We have
\beq
S[g,\Gamma]=\frac{1}{2\kappa}\int d^{n}x \sqrt{-g}\Big( \alpha R+\beta R^{2}+\gamma R \Box R \Big) \label{ActF}
\eeq
where $\alpha$ is dimensionless\footnote{Of course, the General Relativity limit is recovered for $\alpha=1$.} and $\beta$ and $\gamma$ are parameters with mass dimension $-2$ and $-4$ respectively. Also, we define the box operator associated to the metric $g$, in the usual way, (see eq.
(\ref{Boxdef})) as we explained in the previous section. Let us also note that, in this action, the scalar curvature $R$ is now the generic one that  has also torsion and non-metricity.
The action (\ref{ActF}) is, up to surface terms, equivalent to\footnote{\textcolor{black}{Of course when acting all scalars, the covariant derivative, of any connection, always reduces to the partial one \cite{schouten1954ricci}. Therefore, for the scalar curvature we have $\nabla_{\mu}R=\widetilde{\nabla}_{\mu}R=\partial_{\mu}R$}.}
\beq
S[g,\Gamma]=\frac{1}{2\kappa}\int d^{n}x \sqrt{-g}\Big( \alpha R+\beta R^{2}-\gamma g^{\mu\nu} \partial_{\mu}R  \partial_{\nu}R \Big)\,.
\eeq
We shall use the latter form since it is more convenient to perform the variations. Varying with respect to the metric and the independent affine-connection, we obtain the field equations of the non-local theory
\beq
-\frac{1}{2}\Big( \alpha R +\beta R^{2}-\gamma (\partial R)^{2}\Big)g_{\mu\nu} +\Big( \alpha +2 \beta R
+2 \gamma \Box R \Big) R_{(\mu\nu)}-\gamma \partial_{\mu}R \partial_{\nu}R=0 \label{mfesimple}
\eeq
and
\beq
\phi P_{\lambda}{}^{\mu\nu}+\delta_{\lambda}^{\nu}\partial^{\mu}\phi - g^{\mu\nu}\partial_{\lambda}\phi=0 \label{confeP}
\eeq
respectively, where
\begin{gather}
	P_{\lambda}^{\;\;\;\mu\nu}=-\frac{\nabla_{\lambda}(\sqrt{-g}g^{\mu\nu})}{\sqrt{-g}}+\frac{\nabla_{\sigma}(\sqrt{-g}g^{\mu\sigma})\delta^{\nu}_{\lambda}}{\sqrt{-g}} 
	+2(S_{\lambda}g^{\mu\nu}-S^{\mu}\delta_{\lambda}^{\nu}+g^{\mu\sigma}S_{\sigma\lambda}^{\;\;\;\;\nu})  \nonumber \\
=\delta^{\nu}_{\lambda}\left( q^{\mu}-\frac{1}{2}Q^{\mu}-2 S^{\mu} \right) + g^{\mu\nu}\left( \frac{1}{2}Q_{\lambda}+2 S_{\lambda} \right)-( Q_{\lambda}^{\;\;\;\mu\nu}+2 S_{\lambda}^{\;\;\;\;\mu\nu}) \label{Palatini}
	\end{gather}
is the so-called Palatini tensor and we have abbreviated
\beq
(\partial R)^{2}=g^{\mu\nu}\partial_{\mu}R\partial_{\nu}R \;\;, \;\; \phi=\alpha+2\beta R+2 \gamma \Box R
\eeq
Using the method developed in \cite{Iosifidis:2018jwu}, from the above connection field equations, one finds the distortion
\beq
N_{\alpha\mu\nu}=\frac{2}{(n-2)\phi}g_{\nu[\alpha}\partial_{\mu]}\phi+\frac{1}{2}g_{\alpha\mu}q_{\nu}
\eeq
\textcolor{black}{and consequently, torsion and non-metricity read
\beq
S_{\mu\nu}{}{}^{\lambda}=\delta^{\lambda}_{[\mu}\left( \frac{1}{2}q_{\nu]}-\frac{1}{(n-2)}\frac{\partial_{\nu]}\phi}{\phi}\right)
\eeq
\beq
Q_{\nu\alpha\mu}=\frac{1}{2}g_{\alpha\mu}q_{\nu}
\eeq
Let us observe that both are of purely vectorial type.
}
    Now notice that our action (\ref{ActF}) is invariant under projective transformations of the connection
\beq
\Gamma^{\lambda}{}_{\mu\nu}\mapsto \Gamma^{\lambda}{}_{\mu\nu}+\delta^{\lambda}_{\mu}\xi_{\nu}
\eeq
since it is built solely from $R$ which is itself projectively invariant. This means that there is an undetermined vectorial degree of freedom which can be fixed by a gauge choice \cite{Iosifidis:2018jwu}\footnote{\textcolor{black}{It should be noted that this invariance has implications when matter is added. In particular, it forces the dilation part of the hypermomentum to be vanishing (see for instance \cite{hehl1981metric}). Of course this constraint is avoided if the matter does not couple to the connection but only to the metric.}}. Two special gauge choices are of interest here. The first one is to fix
\beq
\xi_{\nu}=-\frac{1}{2}q_{\nu}
\eeq
in which case, as can be easily seen from (\ref{QSN}), the full non-metricity vanishes
\beq
Q_{\alpha\mu\nu}=0
\eeq
and torsion is given by
\beq
S_{\mu\nu\alpha}=\frac{1}{(n-2)\phi}g_{\alpha[\nu}\partial_{\mu]}\phi
\eeq
We call this the zero non-metricity gauge. On the other hand, if we fix the gauge as
\beq
\xi_{\nu}=-\frac{1}{2}q_{\nu}+\frac{1}{(n-2)}\frac{\partial_{\nu}\phi}{\phi}
\eeq
then the full torsion is found to be zero, i.e.
\beq
S_{\mu\nu\alpha}=0
\eeq
and the non-metricity takes the simple Weyl form
\beq
Q_{\alpha\mu\nu}=\frac{1}{(n-2)\phi}(\partial_{\alpha}\phi)g_{\mu\nu}
\eeq
This is the vanishing torsion gauge. We see therefore a torsion/non-metricity duality similar to the one that appears in $f(R)$ theories \cite{Iosifidis:2018zjj}. Let us now investigate the metric field equations (\ref{mfesimple}). Taking the trace of the latter, it follows that
\beq
\left( \frac{n}{2}-1\right) \Big( \gamma (\partial R)^{2}-\alpha R\Big)+2 \beta \left( 1-\frac{n}{4}\right) R^{2}+2 \gamma R \Box R=0 \label{treq}
\eeq
The last describes the propagation of a scalar mode.
From now on let us set $\alpha=1$, $\beta=0$ and $n=4$ and also denote the scalar mode as $R\equiv \psi$. Substituting the trace eq. (\ref{treq}) back in (\ref{mfesimple}), the latter simplifies to
\beq
R_{(\mu\nu)}-\frac{R}{2}g_{\mu\nu}=\frac{1}{\phi}\left( \gamma \partial_{\mu}\psi\partial_{\nu}\psi-\frac{\psi}{2}g_{\mu\nu}\right)
\eeq
Performing a post-Riemannian expansion on the quantities of the left-hand side of the above, the latter takes the form (see appendix for details)
\beq
\phi \left( \tilde{R}_{\mu\nu}-\frac{\tilde{R}}{2}g_{\mu\nu}\right)=\tilde{\nabla}_{\mu}\tilde{\nabla}_{\nu}\phi-\frac{\partial_{\mu}\phi \partial_{\nu}\phi}{\phi}-\Big[ \Box \phi-\frac{3}{4}\frac{(\partial \phi)^{2}}{\phi}\Big] g_{\mu\nu}+\gamma \partial_{\mu}\psi\partial_{\nu}\psi-\frac{\psi}{2}g_{\mu\nu} \label{resultmfe}
\eeq
with $\phi=1+2\gamma \Box \psi$. We see therefore that the initial theory is on-shell equivalent to a Scalar-Tensor theory with higher order derivatives for the scalar field. It is important to stress that, in contrast to the Riemannian case where the resulting metric field equations contain sixth order derivatives of the scalar field, in the metric-affine approach developed here, we have up to fourth-order derivatives of the scalar field. In order words, with respect to the metric formalism developed in \cite{Gottlober:1989ww}\footnote{See also \cite{Amendola:1993bg} for a cosmological application.}, the metric-affine approach reduces the order of the derivatives describing the dynamics of the extra scalar field.

\section{Non-local Metric-Affine Theories}

\subsection{Nonlocal Ultraviolet $\mathcal{L}=R+RF(\Box)R$ Theories}
We shall now generalize the results of the previous section and consider the Non-local Metric-Affine Gravity given by the action
\beq
S[g,\Gamma]=\frac{1}{2\kappa}\int d^{n}x \sqrt{-g}\Big( R+R F(\Box)R \Big) \label{F}
\eeq
where $F(\Box)$ is an analytic function of the box operator, usually called the form factor \cite{BasiBeneito:2022wux}, which we may expand as
\beq
F(\Box)=\sum_{m=1}^{\infty}\frac{a_{m}}{M^{2m}}\Box^{m}=\sum_{m=1}^{\infty}f_{m}\Box^{m}\,.
\eeq
In the first representation $M^{2}$ carries a squared mass scale and consequently the coefficients $a_{m}$ are dimensionless constants. In the second form, the mass dependence is  absorbed in $f_{m}$. In order to have more simplified expressions, we shall use the latter representation. Non-local theories of this kind, namely consisting of positive powers of the box operator are relevant in the UV regime \cite{Buoninfante:2022trw}. In the metric formalism,  it was shown in \cite{Biswas:2005qr} that the class (\ref{F})  gives rise to ghost and asymptotically free theories. Below we shall analyze the dynamics of these theories in the metric-affine approach.

Let us firstly derive the field equations associated with (\ref{F}). Since this is the first time that non-local theories are studied in the Metric-Affine Gravity (MAG) framework, we will carefully provide some  technical details that are needed for the derivation of the field equations. We will start with the metric variations which are the most elaborate. We start with the standard and trivial results
\beq
\delta_{g}\sqrt{-g}=-\frac{1}{2}\sqrt{-g}g_{\mu\nu}\delta g^{\mu\nu}
\eeq
\beq
\delta_{g}R=(\delta g^{\mu\nu})R_{(\mu\nu)}
\eeq
Next, as a first step in dealing  with the $\sqrt{-g}R(\delta \Box^{m}) R$ term, that will appear after expanding $F$, we  recall the identity
\beq
\delta(\Box)\phi=\delta(\Box \phi)-\Box(\delta \phi)
\eeq
for any variation-$\delta$ and for an arbitrary scalar field $\phi$. Given that $\Box \phi=g^{\mu\nu}\widetilde{\nabla}_{\mu}\widetilde{\nabla}_{\nu}$, it immediately follows that
\beq
\delta_{g}(\Box)\phi=(\delta g^{\mu\nu}) \widetilde{\nabla}_{\mu}\widetilde{\nabla}_{\nu}\phi-g^{\mu\nu}(\delta_{g}\tilde{\Gamma}^{\lambda}{}_{\mu\nu})\partial_{\lambda}\phi
\eeq
Then, varying the Levi-Civita connection, we readily obtain
\beq
\delta_{g}\tilde{\Gamma}^{\lambda}{}_{\mu\nu}=\frac{1}{2}g^{\alpha\lambda}\Big( \widetilde{\nabla}_{\mu}(\delta g_{\nu\alpha})+\widetilde{\nabla}_{\nu}(\delta g_{\mu\alpha})-\widetilde{\nabla}_{\alpha}(\delta g_{\mu\nu})\Big) \label{deltagGamma}
\eeq
Using the above two, after some partial integrations and using the identity $\delta g_{\alpha\beta}=-g_{\mu\alpha}g_{\nu\beta}\delta g^{\mu\nu}$, we find
\beq
\sqrt{-g}f (\delta_{g}\Box) \phi=\sqrt{-g}(\delta g^{\mu\nu})\left( -(\widetilde{\nabla}_{(\mu}f) (\widetilde{\nabla}_{\nu)}\phi)+ \frac{f}{2}(\Box \phi)g_{\mu\nu}+\frac{1}{2}(\widetilde{\nabla}_{\alpha}f)(\widetilde{\nabla}^{\alpha}\phi)g_{\mu\nu} \right)+t.d. \label{fdg}
\eeq
up to total derivative terms $t.d.$ which we shall be dropping from now on. Note also the cancellation of the double derivative term $f\widetilde{\nabla}_{\mu}\widetilde{\nabla}_{\nu}\phi$ from the final expression. 
Now we have everything needed to tackle the variation of the $\sqrt{-g}R F(\Box)R$ term. Expressing the form factor in the power-series mentioned above, we have
\beq
\delta_{g}(\sqrt{-g}R F(\Box)R)=\sum_{m=1}^{\infty} f_{m}\delta_{g}(\sqrt{-g}R\Box^{m}R) \label{drfr}
\eeq
we isolate
\beq
\delta_{g}(\sqrt{-g}R\Box^{m}R)=-\frac{1}{2}\sqrt{-g}R\Box^{m}R g_{\mu\nu}(\delta g^{\mu\nu})+\sqrt{-g}(\delta_{g}R) \Box^{m}R+\sqrt{-g}R (\delta \Box^{m})R+\sqrt{-g}R\Box^{m}(\delta R)
\eeq
then, using the fact that up to total derivatives $\sqrt{-g}f\Box \phi=\sqrt{-g}(\Box f) \phi$ if we apply this m-times to the last term on the right-hand side of the above, we see that it is equal to the second. Collecting them and using  $\delta_{g}R=(\delta g^{\mu\nu})R_{(\mu\nu)}$  it follows that
\beq
\delta_{g}(\sqrt{-g}R\Box^{m}R)=-\frac{1}{2}\sqrt{-g}R\Box^{m}R g_{\mu\nu}(\delta g^{\mu\nu})+\sqrt{-g}2 R_{(\mu\nu)}(\Box^{m}R) (\delta g^{\mu\nu}) +\sqrt{-g}R (\delta \Box^{m})R \label{lpart}\,.
\eeq
Now only the last variation remains. In order to compute it, we use the identity
\beq
\delta \Box^{m}=\sum_{l=0}^{m-1}\Box^{l}(\delta \Box) \Box^{m-l-1} \label{varboxmult}
\eeq
along with (\ref{fdg}), to which we set $f=\Box^{l}R=R^{[l]}$ and $\phi=\Box^{m-l-1}R=R^{[m-l-1]}$, to arrive at
\begin{gather}
    \sqrt{-g}R (\delta \Box^{m})R=\sum_{l=0}^{m-1} \sqrt{-g} R \Box^{l}(\delta_{g}\Box)\Box^{m-l-1}R=\sum_{l=0}^{m-1} \sqrt{-g} ( \Box^{l}R) (\delta_{g}\Box)\Box^{m-l-1}R= \nonumber \\
    =\sqrt{-g}(\delta g^{\mu\nu})  \sum_{l=0}^{m-1}\left( -(\widetilde{\nabla}_{(\mu}R^{[l]}) (\widetilde{\nabla}_{\nu)}R^{[m-l-1]})+ \frac{1}{2} R^{[l]}R^{[m-l]}g_{\mu\nu}+\frac{1}{2}(\widetilde{\nabla}_{\alpha}R^{[l]})(\widetilde{\nabla}^{\alpha}R^{[m-l-1]})g_{\mu\nu} \right)
\end{gather}
Substituting the latter into (\ref{lpart}),   we obtain 
\begin{gather}
    \delta_{g}(\sqrt{-g}R\Box^{m}R)=\sqrt{-g}(\delta g^{\mu\nu})\Big[ -\frac{1}{2}R(\Box^{m}R) g_{\mu\nu}+2 R_{(\mu\nu)}\Box^{m}R \nonumber \\+\sum_{l=0}^{m-1}\Big( -(\widetilde{\nabla}_{(\mu}R^{[l]}) (\widetilde{\nabla}_{\nu)}R^{[m-l-1]})+ \frac{1}{2} R^{[l]}R^{[m-l]}g_{\mu\nu}+\frac{1}{2}(\widetilde{\nabla}_{\alpha}R^{[l]})(\widetilde{\nabla}^{\alpha}R^{[m-l-1]})g_{\mu\nu} \Big) \Big]
\end{gather}
Taking the $l=0$ value separately  and consequently starting the sum from $l=1$ the first term in the parenthesis is canceled and we arrive at the final form
\begin{gather}
    \delta_{g}(\sqrt{-g}R\Box^{m}R)=\sqrt{-g}(\delta g^{\mu\nu})\Big[ 2 R_{(\mu\nu)}\Box^{m}R -(\widetilde{\nabla}_{(\mu}R)(\widetilde{\nabla}_{\nu)}R^{[m-1]}) +\frac{1}{2}(\widetilde{\nabla}_{\alpha}R) (\widetilde{\nabla}^{\alpha}R^{[m-1]})g_{\mu\nu}    \nonumber \\+\sum_{l=1}^{m-1}\Big( -(\widetilde{\nabla}_{(\mu}R^{[l]}) (\widetilde{\nabla}_{\nu)}R^{[m-l-1]})+ \frac{1}{2} R^{[l]}R^{[m-l]}g_{\mu\nu}+\frac{1}{2}(\widetilde{\nabla}_{\alpha}R^{[l]})(\widetilde{\nabla}^{\alpha}R^{[m-l-1]})g_{\mu\nu} \Big) \Big]
\end{gather}

With this we are now in a position to derive the metric field equations corresponding to (\ref{ActF}), which, after using all the above results, read
\begin{gather}
R_{(\mu\nu)}-\frac{R}{2}g_{\mu\nu}+2 R_{(\mu\nu)}F(\Box)R-(\widetilde{\nabla}_{(\mu}R)(\widetilde{\nabla}_{\nu)}F(\Box)\Box^{-1}R) +\frac{1}{2}(\widetilde{\nabla}_{\alpha}R) (\widetilde{\nabla}^{\alpha}F(\Box)\Box^{-1}R)g_{\mu\nu}   \nonumber \\
+\sum_{m=2}^{\infty}\sum_{l=1}^{m-1}f_{m}\Big( -(\widetilde{\nabla}_{(\mu}R^{[l]}) (\widetilde{\nabla}_{\nu)}R^{[m-l-1]})+ \frac{1}{2} R^{[l]}R^{[m-l]}g_{\mu\nu}+\frac{1}{2}(\widetilde{\nabla}_{\alpha}R^{[l]})(\widetilde{\nabla}^{\alpha}R^{[m-l-1]})g_{\mu\nu} \Big) \Big]=0
\end{gather}
where $\Box^{-1}$ is the inverse operator of $\Box$, i.e. $\Box \Box^{-1}=\Box^{-1}\Box=\mathbb{1}$ and therefore $F(\Box)\Box^{-1}=\sum_{m=1}^{\infty}f_{m}\Box^{m-1}=f_{1}+f_{2}\Box+...$ namely the sum start from a constant term. As a crosscheck, for $f_{1}=\gamma$, $f_{m}=0$, $m\geq 2$ it can be easily seen that the latter field equations are consistent with (\ref{mfesimple}). Let us now proceed with the connection field equations. These are quite trivial since 
\beq
\delta_{\Gamma}(\sqrt{-g}R\Box^{m}R)=\sqrt{-g}(\delta_{\Gamma}R)\Box^{m}R+\sqrt{-g}R\Box^{m}(\delta_{\Gamma}R)=2\sqrt{-g}(\delta_{\Gamma}R)\Box^{m}R+t.d.
\eeq
where in the last equality we used the fact that $\sqrt{-g}f \Box^{m}\phi=\sqrt{-g}\phi \Box^{m}f+t.d.$, with $t.d.$ denoting total derivative terms. Then, the connection field equations are found to be
\beq
\phi P_{\lambda}{}^{\mu\nu}+\delta_{\lambda}^{\nu}\partial^{\mu}\phi - g^{\mu\nu}\partial_{\lambda}\phi=0 \label{ppjk}
\eeq
where $P_{\lambda}{}^{\mu\nu}$ is the Palatini tensor defined in (\ref{Palatini}) and
\beq
\phi=1+2 F(\Box)R \label{phiF}
\eeq
Note that these are formally the same with (\ref{confeP}), with the mere replacement $\gamma \Box R \rightarrow F(\Box)R $ and of course setting $\alpha=1$ and $\beta=0$. It is therefore again possible to solve for the connection. Indeed, 
despite the complexity of the metric field equations, the connection field equations obtained above, have a  rather simple structure. As a matter of fact, since the form of (\ref{ppjk}) is identical to that of (\ref{confeP}) with only the exact form of $\phi$ altered, the \textcolor{black}{solution}  to (\ref{ppjk}) is still 
\beq
N_{\alpha\mu\nu}=\frac{2}{(n-2)\phi}g_{\nu[\alpha}\partial_{\mu]}\phi+\frac{1}{2}g_{\alpha\mu}q_{\nu}
\eeq
with $\phi$ given by eq. (\ref{phiF}). One can then proceed in a similar way to that of the previous section and bring the metric field equations in the form of Scalar-Tensor theories with higher order derivative terms for the scalar field.

\subsection{Non-local Infrared $\mathcal{L}=R+RF(\Box^{-1})R$ Theories}

The previous results can be straightforwardly applied to non-local theories involving the inverse box operator \cite{Deser:2007jk} $\Box^{-1}$ defined through $\Box \Box^{-1}=\Box^{-1}\Box=\mathbb{1}$.  As a prototype let us consider the class of non-local theories given by\footnote{For the study of such theories in the metric formalism see e.g. \cite{Capozziello:2020xem,Capozziello:2021bki,Capozziello:2023mju}.} 
\beq
S[g,\Gamma]=\frac{1}{2\kappa}\int d^{n}x \sqrt{-g}\Big( R+R F(\Box^{-1})R \Big) \label{F}
\eeq
where $F(\Box^{-1})$ is now an analytic function of the  inverse box operator, 
 which is expanded as
\beq
F(\Box^{-1})=\sum_{m=1}^{\infty}c_{m}\Box^{-m}
\eeq
where $\Box^{-m}$ is an abbreviation for $(\Box^{-1})^{m}$. Note that this form of series for $F$, namely containing powers of the inverse box operator are relevant for the IR regime. In order to proceed with the field equations, we only need to vary the defining relation of $\Box^{-1}$  and subsequently multiply from the left by the same operator, to trivially get\footnote{\textcolor{black}{We assume that a Green’s function and
boundary conditions are fixed and no homogeneous modes of  $\Box$ appear. }}
\beq
\delta \Box^{-1}=-\Box^{-1}(\delta \Box)\Box^{-1}
\eeq
With this result and using those  of the previous subsection, varying with respect to the metric and the connection, we obtain the field equations
\begin{gather}
R_{(\mu\nu)}-\frac{R}{2}g_{\mu\nu}+2 R_{(\mu\nu)}F(\Box^{-1})R-\frac{g_{\mu\nu}}{2}RF(\Box^{-1})R \\
+\sum_{m=1}^{\infty}\sum_{l=0}^{m-1}c_{m}\Big( -(\widetilde{\nabla}_{(\mu}R^{-[l+1]}) (\widetilde{\nabla}_{\nu)}R^{-[m-l]})+ \frac{1}{2} R^{-[l+1]}R^{-[m-l-1]}g_{\mu\nu}+\frac{1}{2}(\widetilde{\nabla}_{\alpha}R^{-[l+1]})(\widetilde{\nabla}^{\alpha}R^{-[m-l]})g_{\mu\nu} \Big) \Big]=0
\end{gather}
and
\beq
\Omega P_{\lambda}{}^{\mu\nu}+\delta_{\lambda}^{\nu}\partial^{\mu}\Omega - g^{\mu\nu}\partial_{\lambda}\Omega=0 \;\;, \;\;
\Omega=1+2 F(\Box^{-1})R \label{phiF}
\eeq
respectively. In the above we have used the notation $R^{-[m]}:=\Box^{-m}R=(\Box^{-1})^{m}R$. It is worth stressing  again that  the connection field equations are formally the same with (\ref{confeP}), with $\phi$ replaced by $\Omega$. Finally, it goes without saying that one can combine the above results and derive the field equations for theories that contain both the box operator and its inverse at the same time. Let us now move on and develop some novel Non-local Metric-Affine Gravity theories which have no Riemannian analogue.

\section{Novel Theories with no Riemannian analogue}

As we have already pointed out, the action (\ref{ActF}) as well as its extension (\ref{F}), have been well studied in the Riemannian case. In particular, in a Riemannian setting, (\ref{ActF}) is the most general theory that is quadratic in scalar curvature. We note, however, that in MAG, squared torsion and non-metricity terms supplement the linear in curvature Einstein-Hilbert term. Therefore, in a MAG scheme before extending to (\ref{ActF}) it would be quite natural to consider non-local actions consisting of scalars formed by sandwiching the box operator with torsion and non-metricity before moving to quadratic curvature.  Given that there are 11 parity even quadratic invariants \textcolor{black}{(see \cite{Pagani:2015ema} and also \cite{iosifidis2019scale} for the proof that this is the number of independent invariants), we get the following  scalars
\newline
\\
	\textbf{Pure Non-Metricity Scalars}
	\begin{gather}
	A_{1}=Q_{\alpha\mu\nu}Q^{\alpha\mu\nu} \\
	A_{2}=Q_{\alpha\mu\nu}Q^{\mu\nu\alpha} \\
	A_{3}=Q_{\mu}Q^{\mu}
	\\
	A_{4}=q_{\mu}q^{\mu}
	\\
	A_{5}=Q_{\mu}q^{\mu}
	\end{gather}
	\newline
	\textbf{Pure Torsion Scalars}
	\begin{gather}
	B_{1}=S_{\alpha\mu\nu}S^{\alpha\mu\nu} \\
	B_{2}=S_{\alpha\mu\nu}S^{\mu\nu\alpha} \\
	B_{3}=S_{\mu}S^{\mu}
	\end{gather}
\newline
	\textbf{Mixed}
	\begin{gather}
	C_{1}=Q_{\alpha\mu\nu}S^{\alpha\mu\nu}
	\\
	C_{2}=Q_{\mu}S^{\mu} 
	\\
	C_{3}=q_{\mu}S^{\mu}\,,
	\end{gather}} 
    
and then the analogue of the theory  (\ref{ActF})   to MAG is
\begin{gather}
S_[g, \Gamma]=\frac{1}{2 \kappa}\int d^{n}x \sqrt{-g}\Big[ R+ b_{1}S_{\alpha\mu\nu}S^{\alpha\mu\nu} +
b_{2}S_{\alpha\mu\nu}S^{\mu\nu\alpha} +
b_{3}S_{\mu}S^{\mu} 
+c_{1}Q_{\alpha\mu\nu}S^{\alpha\mu\nu}+
c_{2}Q_{\mu}S^{\mu} +
c_{3}q_{\mu}S^{\mu} \nonumber \\
a_{1}Q_{\alpha\mu\nu}Q^{\alpha\mu\nu} +
a_{2}Q_{\alpha\mu\nu}Q^{\mu\nu\alpha} +
a_{3}Q_{\mu}Q^{\mu}+
a_{4}q_{\mu}q^{\mu}+
a_{5}Q_{\mu}q^{\mu} 	\nonumber \\
b'_{1}S_{\alpha\mu\nu}\Box S^{\alpha\mu\nu} +
b'_{2}S_{\alpha\mu\nu}\Box S^{\mu\nu\alpha} +
b'_{3}S_{\mu}\Box S^{\mu} 
+c'_{1}Q_{\alpha\mu\nu}\Box S^{\alpha\mu\nu}+
c'_{2}Q_{\mu} \Box S^{\mu} +
c'_{3}q_{\mu}\Box S^{\mu} \nonumber \\
a'_{1}Q_{\alpha\mu\nu}\Box Q^{\alpha\mu\nu} +
a'_{2}Q_{\alpha\mu\nu} \Box Q^{\mu\nu\alpha} +
a'_{3}Q_{\mu}\Box Q^{\mu}+
a'_{4}q_{\mu}\Box q^{\mu}+
a'_{5}Q_{\mu}\Box q^{\mu} \Big] \label{Sboxgen}
\end{gather}
Some comments are in order. The first two lines consist of the usual Einstein-Hilbert term along with the quadratic invariants in torsion and non-metricity. These exhaust the number of parity even invariants up to dimension $[L^{-2}]$. When only the first two lines of the above action are considered then the vacuum theory reduces to General Relativity \cite{Iosifidis:2021tvx}. The third and fourth lines are the new non-local terms we consider. These have not been studied previously in the literature.  Let  qualitatively study their dynamics. The full field equations are quite lengthy and are given in the appendix. Suppressing indices, the connection field equations schematically look like
\beq
\Box N \propto N
\eeq
where $N$ is the distortion. This shows that torsion and non-metricity \textcolor{black}{have dynamics in this case and the theory has more degrees of freedom than General Relativity. It goes beyond the scope of the current study to investigate the healthy sub-sectors of this class of theories}. However,  it is worth mentioning that the non-local terms of the third and fourth lines, when added to the quadratic curvature MAG theories, may help to extend the healthy spectrum of the quadratic MAG. 
In order to get our point across, let us study a  simple, but yet non-trivial, subsector of the above class. 

\subsection{An Example}

Let us study the model given by 
\begin{gather}
S_[g, \Gamma]=\frac{1}{2 \kappa}\int d^{n}x \sqrt{-g}\Big[ R+ \alpha S_{\mu}S^{\mu} +
\beta S_{\mu}\Box S^{\mu}  \Big] \label{SboxS}
\end{gather}
which corresponds to setting all constants in (\ref{Sboxgen}) to zero apart from $b_{3}=\alpha$ and $b_{3}'=\beta$. Varying with respect to the connection, we obtain
\beq
P_{\lambda}{}^{\mu\nu}+2(\alpha S^{[\mu}+\beta \Box S^{[\mu})\delta^{\nu]}_{\lambda}=0 \label{PSBS}
\eeq
Tracing in $\lambda=\mu$ and using the fact that $P_{\mu}{}^{\mu\nu}=0$, it follows that
\beq
\alpha S_{\mu}+\beta \Box S_{\mu}=0 \label{dynS}
\eeq
which is a dynamical equation for the torsion vector. Therefore, in this case, torsion propagates in vacuum due to the inclusion of the non-local term.  Substituting this back to (\ref{PSBS}) gives $P_{\lambda}{}^{\mu\nu}=0$ and consequently the distortion is readily computed to be
\beq
N_{\alpha\mu\nu}=\frac{1}{2}g_{\alpha\mu}q_{\nu} \label{Nform}\,.
\eeq
With this result, we then find the full forms of torsion and non-metricity
\beq
S_{\mu\nu\alpha}=\frac{2}{(n-1)}S_{[\mu}g_{\nu]\alpha} \;\;, \;\; Q_{\alpha\mu\nu}=\frac{Q_{\alpha}}{n}g_{\mu\nu}
\eeq
where the vectors are related through
\beq
Q_{\mu}=n q_{\mu}=-\frac{4 n}{(n-1)}S_{\mu}
\eeq
and the whole dynamics is contained in (\ref{dynS}). It has to be noted that the distortion (\ref{Nform}) is of a very peculiar type (purely projective mode). It has the remarkable property to give no post-Riemannian contributions to the symmetric part of the Ricci tensor and, consequently, also the Ricci scalar, is
\beq
R_{(\mu\nu)}=\tilde{R}_{\mu\nu}\;\;, \;\; R=\tilde{R}\,.
\eeq
This is of course tightly related with the fact that  two connections  are related through a projective transformation and have their symmetric Ricci tensor equal. Furthermore the connections, in this equivalence class, share also the same autoparallel curves up to a reparameterization of the affine parameter \cite{Iosifidis:2018zjj}.

Finally, varying with respect to the connection and using the form of (\ref{Nform}) we find the metric field equations
\begin{gather}
    \tilde{R}_{\mu\nu}-\frac{1}{2}\tilde{R} g_{\mu\nu}=\frac{\alpha}{2}S_{\alpha}S^{\alpha}g_{\mu\nu}-\alpha S_{\mu}S_{\nu}-\frac{\beta}{2}(\widetilde{\nabla}_{\alpha}S^{\alpha})^{2}g_{\mu\nu} \nonumber \\
    +\beta \Big( \widetilde{\nabla}_{\mu}S_{\alpha}\widetilde{\nabla}_{\nu}S^{\alpha}+\widetilde{\nabla}_{\alpha}S_{\mu}\widetilde{\nabla}^{\alpha}S_{\nu}\Big) -\frac{\beta}{2}\widetilde{\nabla}^{\alpha}\Big( S_{(\mu}\widetilde{\nabla}_{\nu)}S_{\alpha}+(\widetilde{\nabla}_{\alpha}S_{(\mu})S_{\nu)}-S_{\alpha}\widetilde{\nabla}_{(\mu}S_{\nu)}\Big)\label{RSfe}
\end{gather}
from which it is clear that the resulting theory is a Vector-Tensor one, where the vector is of geometric origin. Of course, it goes without saying that in general, these excitations extra degrees of freedom will lead to pathologies. However, it should be noted that the additional terms we consider here when added to the quadratic MAG theory could extend the spectrum of healthy theories just like the inclusions of cubic terms in the action, see \cite{Bahamonde:2024efl}.

\section{Applications to Cosmology}

Let us now discuss some applications for our development. Consider a flat Fiedmann-Lemaìtre-Roberston-Walker (FLRW) Universe with the usual Robertson-Walker line element in Cartesian coordinates $ds^{2}=-dt^{2}+a^{2}(t)(dx^{2}+dy^{2}+dz^{2})$, where $a(t)$ is the scale factor. We shall first derive the modified Friedmann equations for the model (\ref{ActF}). Taking the $00$ and $ij$ components of the resulting field equations (\ref{resultmfe}), we find 
\beq
3 H^{2}=-\frac{1}{4}\left( \frac{\dot{\phi}}{\phi} \right)^{2}-3 H \left( \frac{\dot{\phi}}{\phi} \right)+\frac{1}{\phi}\left( \gamma \dot{\psi}^{2}+\frac{\psi}{2}\right)\,,
\eeq
\beq
\frac{\ddot{a}}{a}=\frac{5}{12}\left(\frac{\dot{\phi}}{\phi} \right)^{2}-\frac{1}{3}H \left(\frac{\dot{\phi}}{\phi}\right)-\frac{\ddot{\phi}}{2 \phi}+\frac{1}{6 \phi}(\psi-\gamma \dot{\psi}^{2})\,,
\eeq
where $\phi=1-6 \gamma (\ddot{\psi}+3 H \dot{\psi})$. In addition, the evolution eq. (\ref{treq}) for the scalar field  takes the form
\beq
2 \gamma \psi (\ddot{\psi}+3 H \dot{\psi})+\psi+\gamma \dot{\psi}^{2}=0 \label{Hpsi}
\eeq
The latter together with the first Friedmann equation above, fully determine the dynamics of the system. Note that for $\gamma=0$, it follows that $\psi=0$ and $\phi=1$ and the above reduce to the usual Friedmann equations as expected. Let us find solutions to the above system. With the premise that non-local effects (attributed to the scalar $\phi$) manifest themselves at the early stages of the Universe history, it is meaningful to consider the ansatz
\beq
\psi=\psi_{0}e^{-b t} \label{psi}
\eeq
with $b>0$. The value of this parameter is obtained by imposing consistency on the system of equations. The above ansatz ensures that non-local quantum-mechanical effects are important at very early times and quickly diminish as we enter into the classical regime. Substituting (\ref{psi}) into (\ref{Hpsi}), it then follows that
\beq
H=\lambda b e^{b t}+\frac{b}{2}  \label{H}
\eeq
where $\lambda=\frac{1}{6 \gamma b^{2}\psi_{0}}$. The latter immediately integrates to
\beq
a(t)=a_{0}e^{{\lambda}e^{bt}+\frac{b}{2}t}
\eeq
namely, non-locality produces a superexponential expansion. Further substitution of (\ref{psi}) and (\ref{H}) fixes the  value of b. This simple application demonstrates the importance of early  non-local effects in non-Riemannian Universes.

For completeness, we will  also report the corresponding cosmological equations for the novel theory (\ref{SboxS}).  Let us recall that, in this case, since only the vector part of torsion is excited, in the FLRW spacetime the full torsion is fully described by a single scalar field and the torsion vector takes the form \cite{tsamparlis1979cosmological}
\beq
S_{\mu}=3 \Phi u_{\mu}
\eeq
Now, contracting \eqref{dynS} with $S^{\mu}$ and using the identity $S^{\mu}\Box S_{\mu}=\frac{1}{2}\Box (S_{\mu}S^{\mu})-\tilde{\nabla}^{\alpha}S^{\beta}\tilde{\nabla}_{\alpha}S_{\beta}$ along with the above form of torsion, we readily obtain the evolution equation
\beq
\alpha \Phi^{2}+\frac{\beta}{6}\Big( \ddot{(\Phi^{2}) }+3 H \dot{(\Phi)^{2}}\Big) +\beta (\dot{\Phi})^{2}+3\beta H^{2}\Phi^{2}=0
\eeq
for the torsion field. Finally, taking the trace of (\ref{RSfe}), we find a variant of the Friedmann equation
\beq
2 \left( H^{2}+\frac{\ddot{a}}{a}\right)=- 3 \alpha \Phi^{2}+\frac{9 \beta}{2}(\dot{\Phi}+3 H \Phi)^{2}-6 \beta (\dot{\Phi}+3 H^{2}\Phi^{2})+3 \beta(\ddot{\Phi}\Phi+\dot{\Phi}^{2}+3 H \Phi \dot{\Phi})-\frac{\beta}{2}\Phi (\ddot{\Phi}+3 \dot{H}\Phi+3 H \dot{\Phi})
\eeq
The latter two equations fully describe the homogeneous cosmological dynamics of the model (\ref{SboxS}). It goes beyond the purpose of the current work to further analyze such cosmological models, however let us note, as seen from these examples, that non-local MAG offers a rich cosmological dynamics.

\section{Discussion and Conclusions}

We have constructed the metric-affine version of Non-local Gravity. Starting with a simple non-local metric-affine model (see eq. (\ref{ActF})), after solving the associated connection field equations, we showed that the theory is on-shell equivalent to a metric Scalar-Tensor Gravity with higher order derivative terms for the propagating scalar field. We then generalized the results to generic Infinite Derivative Gravity theories of the type $\mathcal{L}=RF(\Box)R$, where $R$ is the general scalar curvature and $F(\Box)$ is an analytic function of the box operator. In particular, after devising the general method to obtain the field equations, we again successfully solved the connection field equations and  obtained an on-shell equivalence to some higher derivative Scalar-Tensor Gravity. The same analysis was also performed for non-local theories that are relevant in the IR regime, namely of the form $\mathcal{L}=RF(\Box)R$ where $F(\Box^{-1})$ is now an analytic function of the inverse box operator $\Box^{-1}$, defined through $\Box \Box^{-1}=\Box^{-1}\Box=\mathbb{1}$. It was shown that also in this case the connection field equations can be solved analytically and an equivalence to higher derivative Scalar-Tensor Theories was established in this case as well.

Furthermore, we constructed a novel non-local theory, as given by the action (\ref{Sboxgen}), which has no Riemannian analogue. In particular, our proposed theory consists of non-local quantum modifications to torsion and non-metricity that supplement the usual quadratic (in torsion and non-metricity) MAG. For this novel construction, we discussed how both torsion and non-metricity become dynamical even in vacuum. In general, when all  modes of torsion and non-metricity are excited, the theory is equivalent to some generic Scalar-Vector-Tensor theory. Consequently, in this setting, such generalizations, already existing in the literature and including additional scalar and vector fields, here arise naturally from the enrichment of the underlying geometry. In this sense, the extra fields have a geometric origin, coming from torsion and non-metricity.
Finally, we briefly discussed possible cosmological applications of non-local MAG. The full cosmological analysis of such theories will be studied elsewhere.

\section*{Acknowledgements}
 We thank Friedrich H. Hehl for useful discussions and comments. The authors   acknowledge the support of  Istituto Nazionale di Fisica Nucleare (INFN), Sezione  di Napoli,  {\it Iniziative Specifiche} QGSKY and MOONLIGHT2, and the Istituto Nazionale di Alta Matematica (INdAM),  Gruppo Nazionale di Fisica Matematica (GNFM). 
 This paper is based upon work from COST Action CA21136 {\it Addressing observational tensions in cosmology with systematics  and fundamental physics} (CosmoVerse) supported by COST (European Cooperation in Science and Technology). 

\section{Appendix}
\subsection{Non-Riemannian expansions}
Let us gather here some useful formulae that were used in the various derivations. Firstly, in the vanishing non-metricity gauge, namely for the distortion
\beq
N_{\alpha\mu\nu}=\frac{2}{(n-2)}g_{\nu[\alpha}u_{\mu]}
\eeq
where we have set
\beq
u_{\alpha}:=\frac{\partial_{\alpha}\phi}{\phi}=\partial_{\alpha}\ln{|\phi|} \;, 
\eeq
we compute the non-Riemannian expansions
\beq
R_{(\mu\nu)}=\tilde{R}_{\mu\nu}-\tilde{\nabla}_{(\mu}u_{\nu)}+\frac{1}{(n-2)}u_{\mu}u_{\nu}-\frac{1}{(n-2)}\Big( \tilde{\nabla}_{\alpha}u^{\alpha}+u_{\alpha}u^{\alpha} \Big) g_{\mu\nu}
\eeq
and 
\beq
R=\tilde{R}-\left(\frac{n-1}{n-2}\right)\Big( 2 \tilde{\nabla}_{\alpha}u^{\alpha}+u_{\alpha}u^{\alpha} \Big) 
\eeq
From which it follows that
\beq
R_{(\mu\nu)}-\frac{R}{2}g_{\mu\nu}=\tilde{R}_{\mu\nu}-\frac{\tilde{R}}{2}g_{\mu\nu}-\tilde{\nabla}_{(\mu}u_{\nu)}+\left[ \tilde{\nabla}_{\alpha}u^{\alpha}+\frac{(n-3)}{2(n-2)}u_{\alpha}u^{\alpha} \right]g_{\mu\nu}
\eeq

\subsection{Variations}
Here we provide some details related to the variations, since this is the first time this framework is considered. Let us denote with $\delta_{\chi}$ the variation with respect to the $\chi$ which in our case would be either the metric $g$ or the connection $\Gamma$. The first thing to note is that since our box operator contains only the metric, its $\Gamma$-variation immediately vanishes, viz.
\beq
\delta_{\Gamma}\Box=0
\eeq
With this fact let us proceed to derive the connection field equations for the Theory (\ref{Sboxgen}). The variations for the first two lines in (\ref{Sboxgen}) are well known, see e.g. \cite{Iosifidis:2021tvx}, so we will concentrate on the derivation of the new part. Set
\begin{gather}
\mathcal{L}_{\Box}^{(2)}:= b'_{1}S_{\alpha\mu\nu}\Box S^{\alpha\mu\nu} +
b'_{2}S_{\alpha\mu\nu}\Box S^{\mu\nu\alpha} +
b'_{3}S_{\mu}\Box S^{\mu} 
+c'_{1}Q_{\alpha\mu\nu}\Box S^{\alpha\mu\nu}+
c'_{2}Q_{\mu} \Box S^{\mu} +
c'_{3}q_{\mu}\Box S^{\mu} \nonumber \\
a'_{1}Q_{\alpha\mu\nu}\Box Q^{\alpha\mu\nu} +
a'_{2}Q_{\alpha\mu\nu} \Box Q^{\mu\nu\alpha} +
a'_{3}Q_{\mu}\Box Q^{\mu}+
a'_{4}q_{\mu}\Box q^{\mu}+
a'_{5}Q_{\mu}\Box q^{\mu}
\end{gather}
Then after some partial integrations and relabeling of the indices we can gather
\beq
\delta_{\Gamma}\Big(\sqrt{-g}\mathcal{L}_{\Box}^{(2)}\Big)=(\Box \Sigma^{\alpha\mu\nu})\delta_{\Gamma}S_{\alpha\mu\nu}+(\Box \Omega^{\alpha\mu\nu})\delta_{\Gamma}Q_{\alpha\mu\nu}+t.d.
\eeq
where
\beq
\Sigma^{\alpha\mu\nu}=2 b_{1}'S^{\alpha\mu\nu}+b_{2}'S^{\nu\alpha\mu}+b_{2}'S^{\mu\nu\alpha}+2 b_{3}'S^{\alpha}g^{\mu\nu}+c_{1}'Q^{\alpha\mu\nu}+c_{2}'Q^{\alpha}g^{\mu\nu}+c_{3}'q^{\alpha}g^{\mu\nu}
\eeq
\beq
\Omega^{\alpha\mu\nu}=2 a_{1}'Q^{\alpha\mu\nu}+a_{2}'Q^{\nu\alpha\mu}+a_{2}' Q^{\mu\nu\alpha}+c_{1}' S^{\alpha\mu\nu}+ 2 a_{3}'Q^{\alpha}g^{\mu\nu}+2 a_{4}' q^{\nu} g^{\alpha\mu}+a_{5}' Q^{\nu}g^{\alpha\mu}+a_{5}'q^{\alpha}g^{\mu\nu}+c_{2}'Q^{\alpha}g^{\mu\nu}+c_{3}' S^{\nu}g^{\alpha\mu}
\eeq
Then using the variations $\delta_{\Gamma}S_{\alpha\beta}{}{}^{\lambda}=\delta^{[\mu}_{\alpha}\delta^{\nu]}_{\beta}\delta \Gamma^{\lambda}{}_{\mu\nu}$ and $\delta_{\Gamma}Q_{\rho\alpha\beta}=2 \delta^{\nu}_{\rho}\delta^{\mu}_{(\alpha}g_{\beta)\lambda}\delta \Gamma^{\lambda}{}_{\mu\nu}$ we finally arrive at
\beq
\delta_{\Gamma}\Big(\sqrt{-g}\mathcal{L}_{\Box}^{(2)}\Big)=(\delta \Gamma^{\lambda}{}_{\mu\nu})\Big( \Box (\Sigma^{[\mu\nu]}{}{}_{\lambda}+2 \Omega^{\nu(\mu\alpha)}g_{\alpha\lambda})\Big)
\eeq
Consequently, using also known results for the $\Gamma$-variation of the rest of the terms in (\ref{Sboxgen}), we find the connection field equations
\begin{gather}
\left( \frac{Q_{\lambda}}{2}+2 S_{\lambda}\right)g^{\mu\nu}-Q_{\lambda}^{\;\;\mu\nu}-2 S_{\lambda}^{\;\;\mu\nu}+\left( q^{\mu} -\frac{Q^{\mu}}{2}-2 S^{\mu}\right)\delta_{\lambda}^{\nu}+4 a_{1}Q^{\nu\mu}_{\;\;\;\;\lambda}+2 a_{2}(Q^{\mu\nu}_{\;\;\;\;\lambda}+Q_{\lambda}^{\;\;\;\mu\nu})+2 b_{1}S^{\mu\nu}_{\;\;\;\;\lambda} \nonumber \\
+2 b_{2}S_{\lambda}^{\;\;\;[\mu\nu]}+c_{1}\Big( S^{\nu\mu}_{\;\;\;\;\lambda}-S_{\lambda}^{\;\;\;\nu\mu}+Q^{[\mu\nu]}_{\;\;\;\;\;\lambda}\Big)+\delta_{\lambda}^{\mu}\Big( 4 a_{3}Q^{\nu}+2 a_{5}q^{\nu}+2 c_{2}S^{\nu}\Big)+\delta_{\lambda}^{\nu}\Big(  a_{5}Q^{\mu}+2 a_{4}q^{\mu}+ c_{3}S^{\mu}\Big) \nonumber \\
+g^{\mu\nu}\Big(a_{5} Q_{\lambda}+2 a_{4}q_{\lambda}+c_{3}S_{\lambda} \Big)+\Big( c_{2} Q^{[\mu}+ c_{3}q^{[\mu}+2 b_{3}S^{[\mu}\Big) \delta^{\nu]}_{\lambda} 
 +\Big( \Box (\Sigma^{[\mu\nu]}{}{}_{\lambda}+2 \Omega^{\nu(\mu\alpha)}g_{\alpha\lambda})\Big)=0\label{Gfieldeqs}
\end{gather}
From this and using the relations $S_{\mu\nu\alpha}=N_{\alpha[\mu\nu]}$ and $Q_{\alpha\mu\nu}=2 N_{(\mu\nu)\alpha}$ it is then clear that we have a dynamical equation for the distortion which we schematically express as $\Box N=N$ as stated.Let us now illustrate how one can derive the metric field equations for the action (\ref{ActF}). The $g$-variation first two lines that contain no box operator is well known, see e.g. \cite{Iosifidis:2021bad}. The lines containing the box operator are trickier so let us elaborate on them in detail. We shall perform the variation on the first term $\sqrt{-g}S_{\mu\nu\alpha}\Box S^{\mu\nu\alpha}$ and of course the same procedure must then be followed for the rest of the terms. The first thing to do is to perform a partial integration 
\beq
\sqrt{-g}S_{\alpha\mu\nu}\Box S^{\alpha\mu\nu}=\sqrt{-g}g^{\kappa \lambda}\tilde{\nabla}_{\kappa}S_{\alpha\mu\nu}\tilde{\nabla}_{\lambda}S^{\alpha\mu\nu}+t.d.
\eeq
then extract all g-dependence from the mixed torsion
\beq
\sqrt{-g}S_{\alpha\mu\nu}\Box S^{\alpha\mu\nu}=\sqrt{-g}g^{\kappa \lambda}g_{\gamma\rho}g^{\alpha\delta}g^{\beta\epsilon}\tilde{\nabla}_{\kappa}S_{\alpha\beta}{}{}^{\rho}\tilde{\nabla}_{\lambda}S_{\delta \epsilon}{}{}^{\gamma}+t.d.
\eeq
in order to use the fact that the prototype torsion $S_{\mu\nu}{}{}^{\rho}$ is independent from the metric. Employing the Leibniz rule, we obtain
\begin{gather}
    \delta_{g}(\sqrt{-g}S_{\alpha\mu\nu}\Box S^{\alpha\mu\nu})=\sqrt{-g}(\delta g^{\mu\nu}) \Big[  -\frac{1}{2}g_{\mu\nu}(\tilde{\nabla}_{\lambda}S_{\alpha\beta\gamma})(\tilde{\nabla}^{\lambda}S^{\alpha\beta\gamma})+(\tilde{\nabla}_{\mu}S_{\alpha\beta\gamma})(\tilde{\nabla}_{\nu}S^{\alpha\beta\gamma})\nonumber \\-(\tilde{\nabla}^{\lambda}S_{\alpha\beta\nu})(\tilde{\nabla}^{\lambda}S^{\alpha\beta}{}{}_{\mu})
    +2 (\tilde{\nabla}_{\lambda}S_{\mu\alpha\beta})(\tilde{\nabla}^{\lambda}S_{\nu}{}^{\alpha\beta}) \Big]+2 \sqrt{-g}(\tilde{\nabla}^{\lambda}S^{\alpha\beta}{}{}_{\gamma})\delta_{g}(\tilde{\nabla}_{\lambda}S_{\alpha\beta}{}{}^{\gamma})
\end{gather}
We now isolate the last term. Since $\delta_{g}S_{\alpha\beta}{}{}^{\gamma}=0$ we have that
\beq
\delta_{g}(\tilde{\nabla}_{\lambda}S_{\alpha\beta}{}{}^{\gamma})=-(\delta_{g}\tilde{\Gamma}^{\kappa}{}_{\alpha\lambda})S_{\kappa\beta}{}{}^{\gamma}-(\delta_{g}\tilde{\Gamma}^{\kappa}{}_{\beta\lambda})S_{\alpha\kappa}{}{}^{\gamma}+(\delta_{g}\tilde{\Gamma}^{\gamma}{}_{\kappa\lambda})S_{\alpha\beta}{}{}^{\kappa}
\eeq
Using this, after relabeling indices and gathering  terms it follows that
\beq
2 \sqrt{-g}(\tilde{\nabla}^{\lambda}S^{\alpha\beta}{}{}_{\gamma})\delta_{g}(\tilde{\nabla}_{\lambda}S_{\alpha\beta}{}{}^{\gamma})=2\Big( S_{\kappa\lambda}{}{}^{\alpha}(\tilde{\nabla}^{\beta}S^{\kappa\lambda}{}{}_{\gamma} )-2 S_{\gamma\lambda}{}{}^{\kappa}(\tilde{\nabla}^{\beta}S^{\alpha\lambda}{}{}_{\kappa}) \Big)\sqrt{-g}(\delta_{g}\tilde{\Gamma}^{\gamma}{}_{\alpha\beta})
\eeq
and upon using (\ref{deltagGamma}) we finally find
\begin{gather}
  2 \sqrt{-g}(\tilde{\nabla}^{\lambda}S^{\alpha\beta}{}{}_{\gamma})\delta_{g}(\tilde{\nabla}_{\lambda}S_{\alpha\beta}{}{}^{\gamma})=(\delta g^{\mu\nu})  \sqrt{-g}\tilde{\nabla}^{\lambda}\Big( S_{\alpha\beta(\mu}\tilde{\nabla}_{\nu)}S^{\alpha\beta}{}{}_{\lambda}-2 S_{(\mu}{}{}^{\alpha\beta}\tdnb_{\nu)}S_{\lambda\alpha\beta}+ S_{\alpha\beta(\mu}\tdnb_{|\lambda|}S^{\alpha\beta}{}{}_{\nu)} \nonumber \\
  -2 S_{(\mu}{}{}^{\alpha\beta}\tdnb_{|\lambda|}S_{\nu)\alpha\beta}-S_{\alpha\beta\lambda}\tdnb_{(\mu}S^{\alpha\beta}{}{}_{\nu)}+2 S_{\lambda\alpha\beta}\tdnb_{(\mu}S_{\nu)}{}{}^{\alpha\beta}\Big)
\end{gather}
Consequently, for the full variation we have
\begin{gather}
    \delta_{g}(\sqrt{-g}S_{\alpha\mu\nu}\Box S^{\alpha\mu\nu})=\sqrt{-g}(\delta g^{\mu\nu}) \Big[  -\frac{1}{2}g_{\mu\nu}(\tilde{\nabla}_{\lambda}S_{\alpha\beta\gamma})(\tilde{\nabla}^{\lambda}S^{\alpha\beta\gamma})+(\tilde{\nabla}_{\mu}S_{\alpha\beta\gamma})(\tilde{\nabla}_{\nu}S^{\alpha\beta\gamma})\nonumber \\-(\tilde{\nabla}^{\lambda}S_{\alpha\beta\nu})(\tilde{\nabla}^{\lambda}S^{\alpha\beta}{}{}_{\mu})
    +2 (\tilde{\nabla}_{\lambda}S_{\mu\alpha\beta})(\tilde{\nabla}^{\lambda}S_{\nu}{}^{\alpha\beta}) \tilde{\nabla}^{\lambda}\Big( S_{\alpha\beta(\mu}\tilde{\nabla}_{\nu)}S^{\alpha\beta}{}{}_{\lambda}-2 S_{(\mu}{}{}^{\alpha\beta}\tdnb_{\nu)}S_{\lambda\alpha\beta}+ S_{\alpha\beta(\mu}\tdnb_{|\lambda|}S^{\alpha\beta}{}{}_{\nu)} \nonumber \\
  -2 S_{(\mu}{}{}^{\alpha\beta}\tdnb_{|\lambda|}S_{\nu)\alpha\beta}-S_{\alpha\beta\lambda}\tdnb_{(\mu}S^{\alpha\beta}{}{}_{\nu)}+2 S_{\lambda\alpha\beta}\tdnb_{(\mu}S_{\nu)}{}{}^{\alpha\beta}\Big) \Big]
\end{gather}
The same procedure is then followed for the rest of the terms of this kind.

\bibliographystyle{unsrt}
	\bibliography{ref}

\end{document}